\documentclass{article}
\usepackage{makecell}
\usepackage{float}
\usepackage{amsmath}
\DeclareUnicodeCharacter{2606}{}
\usepackage{multirow}
\usepackage{arxiv}

\usepackage[utf8]{inputenc} 
\usepackage[T1]{fontenc}    
\usepackage{hyperref}       
\usepackage{url}            
\usepackage{booktabs}       
\usepackage{amsfonts}       
\usepackage{nicefrac}       
\usepackage{microtype}      
\usepackage{lipsum}		
\usepackage{graphicx}
\usepackage{natbib}
\usepackage{doi}
\usepackage{threeparttable}

\title{Contested Citations: The Role of Open Access Publications in Wikipedia's Scientific Disputes}

\author{ Puyu Yang \\
	Institute for Logic, Language and Computation (ILLC)\\
	University of Amsterdam\\
    1098XH, Amsterdam, The Netherlands.\\
	\texttt{p.yang2@uva.nl} \\
        \And
	Vincent Traag \\
	Centre for Science and Technology Studies (CWTS)\\
	Leiden University\\
	Rapenburg 70, 2311 EZ, Leiden, The Netherlands. \\
	\texttt{v.a.traag@cwts.leidenuniv.nl} \\
	\And
	Rodrigo Costas \\
	Centre for Science and Technology Studies (CWTS)\\
	Leiden University\\
	Rapenburg 70, 2311 EZ, Leiden, The Netherlands. \\
	\texttt{rcostas@cwts.leidenuniv.nl} \\
        \And
	Giovanni Colavizza\thanks{Giovanni Colavizza is also affiliated at the University of Bologna, Department of Classical and Italian Philology, Italy.} \\
	Department of Communication\\
	University of Copenhagen\\
	Karen Blixens Plads 8, Copenhagen, Denmark. \\
	\texttt{colavizza@hum.ku.dk} \\
}

\renewcommand{\headeright}{}
\renewcommand{\shorttitle}{Contested Citations: The Role of Open Access Publications in Wikipedia's Scientific Disputes}

\hypersetup{
pdftitle={A template for the arxiv style},
pdfsubject={q-bio.NC, q-bio.QM},
pdfauthor={David S.~Hippocampus, Elias D.~Striatum},
pdfkeywords={First keyword, Second keyword, More},
}

\begin{document}
\maketitle

\begin{abstract}

Wikipedia is one of the largest online encyclopedias, which relies on scientific publications as authoritative sources. The increasing prevalence of open access (OA) publishing has expanded the public availability of scientific knowledge; however, its impact on the dynamics of knowledge contestation within collaborative environments such as Wikipedia remains underexplored. To address this gap, we analyze a large-scale dataset that combines Wikipedia edit histories with metadata from scientific publications cited in disputed Wikipedia articles. Our study investigates the characteristics of scientific publications involved in disputes and examines whether OA articles are more likely to be contested than paywalled ones.

We find that scientific disputes on Wikipedia are more frequent in the social sciences and humanities, where topics often involve social values and interpretative variability. Publications with higher citation counts and publications in high-impact journals are more likely to be involved in disputes. OA publications are significantly more likely to be involved in disputes and tend to be contested sooner after publication than paywalled articles. This pattern suggests that increased accessibility accelerates both engagement and scrutiny. The relationship between OA status and dispute involvement also varies across disciplines, reflecting differences in Wikipedia editorial practices and norms.

These findings highlight the dual role of OA in both expanding access to scientific knowledge and increasing its visibility in contexts of public negotiation and debate. This study contributes to a broader understanding of how scientific knowledge is collaboratively constructed and contested on open platforms, offering insights for research on open science, scholarly communication, and digital knowledge governance.

\end{abstract}


\section{Introduction}

Wikipedia has emerged as one of the largest and most accessible online encyclopedias in the digital age, serving millions of users worldwide across a wide range of fields. Its open editing model allows virtually anyone to contribute or modify content, which facilitated rapid growth and continuous updates. However, this openness also introduces challenges, including vandalism, the spread of misinformation, and edit conflicts~\citep{priedhorsky2007creating, kittur2007he, sumi2011edit}. These conflicts often arise when multiple editors repeatedly undo each other's contributions due to disagreements~\citep{kittur2007he, yasseri2012dynamics}. A core principle guiding editorial practice on Wikipedia is the Neutral Point of View\footnote{\url{https://en.wikipedia.org/wiki/Wikipedia:Neutral_point_of_view}} (NPOV), which requires that articles be written fairly, accurately, and supported by verifiable sources. This principle is especially important for articles addressing contentious issues or biographies of living persons. Within this framework, not only Wikipedia articles content but also the citations used to support claims can become central to editorial disputes.

In addition to its encyclopedic function, Wikipedia increasingly serves as a platform for public engagement with scientific knowledge~\citep{jemielniak2016bridging, segev2017temporal, shafee2017academics}. Its commitment to transparency, evident in open revision histories, publicly accessible discussion pages, and community-enforced guidelines, has contributed to its credibility among both general readers and professionals. Unlike academic journals, which often rely on closed peer review processes and involve long publication timelines, Wikipedia allows for immediate updates. Editors can incorporate newly published scientific findings in near real-time. This responsiveness makes Wikipedia a dynamic platform where scientific knowledge is not only disseminated but also interpreted, negotiated, and at times contested in public view~\citep{benjakob2018clockwork, black2008wikipedia}.

Recent research has recognized Wikipedia as a space where scientific disputes become visible~\citep{weltevrede2016platform, wilson2015content}. Editors frequently disagree on how to deal with contested scientific claims, and such disputes are often mediated through community policies concerning verifiability, reliable sourcing, and neutrality~\citep{hara2015social, wyatt2016controversy, steiert2025declaring}. These editorial interactions offer valuable insights into how scientific authority is constructed, reinforced, or challenged within a collaborative and public knowledge environment.

While previous research on Wikipedia conflicts has primarily focused on editorial behavior and content disputes, often using metrics such as edit frequency, revert patterns, and editor activity~\citep{yasseri2012dynamics}, the role of scientific citations within these disputes has received comparatively less attention. Scientific publications, typically regarded as authoritative sources, are frequently cited to support or refute competing interpretations. Examining how these sources are mobilized in editorial conflicts can offer important insights into the relationship between scientific knowledge production and its contestation in digital public settings.

One particularly relevant dimension of this issue concerns the role of open access (OA) publications. OA publications, which are openly accessible to all readers, have become increasingly common and are often associated with greater visibility and citation impact~\citep{bjork2012open, piwowar2018state}. Their accessibility may make them more likely to be cited in contentious discussions, as they are easier to retrieve, read, and assess by a broader range of contributors. This raises the possibility that OA publications are more frequently involved in editorial disputes on Wikipedia. Despite this possible connection, empirical research assessing whether OA publications are disproportionately cited in contentious editing contexts remains limited.

In light of these gaps, this study aims to investigate the role of OA publications in disputes identified within Wikipedia articles. Specifically, we address the following research questions:

\begin{itemize}
    \item \textbf{RQ1}: What are the most important characteristics of scientific publications that are involved in scientific disputes?
    \item \textbf{RQ2}: Are OA publications more likely to be used in scientific disputes compared to paywalled publications?
\end{itemize}

To answer these questions, we construct a comprehensive dataset that integrates Wikipedia edit histories with metadata from Crossref and OpenAlex. This dataset includes 3,514 scientific publications cited across 2,221 identified dispute cases on Wikipedia. We analyze the characteristics of these publications and compare OA and paywalled articles in terms of their involvement in disputes, using multiple indicators of \textit{dispute intensity}. Furthermore, we employ logistic regression models to estimate the likelihood that a scientific publication is cited in a dispute, controlling for factors such as article age, citation count, field of research, and journal type.

By examining the relationship between OA status and the likelihood of involvement in citation-related conflicts, this study offers new insights into how scientific knowledge is mobilized, contested, and negotiated within the digital public sphere. Our findings contribute to the literature on science communication, OA, and the governance of user-generated knowledge platforms such as Wikipedia.

\section{Previous Work}
\label{sec:headings}

\subsection{Wikipedia and its Controversies}

Wikipedia is widely recognized as a pioneering open knowledge platform, characterized by its collaborative and open editing model~\citep{kittur2007he}. While this model fosters inclusiveness and facilitates the rapid dissemination of knowledge, it also creates potential for editorial disagreements. One prominent form of conflict is known as an \textit{edit war}, where contributors repeatedly override each other's edits due to differing perspectives~\citep{yasseri2012dynamics}. These conflicts raise important questions about the reliability and neutrality of collaboratively produced content~\citep{arazy2011information}.

In most cases, Wikipedia articles evolve through a predominantly constructive and cooperative process. Editors tend to improve each other's contributions by expanding content, correcting factual or grammatical errors, and gradually converging toward a stable, consensus-based version~\citep{wilkinson2007assessing}. According to \citeauthor{wilkinson2007assessing}, nearly 99\% of articles on the English Wikipedia develop through this relatively smooth and incremental editorial dynamic. Representative examples of such evolution include entries like \texttt{Benjamin Franklin}, \texttt{Pumpkin}, and \texttt{Helium}, which have seen steady development without major conflict.

However, this collaborative ideal does not always hold. In articles dealing with controversial, politicized, or high-profile topics, editorial collaboration can give way to conflict. Disputes often take the form of so-called \textit{edit wars}\footnote{ \url{https://en.wikipedia.org/wiki/Wikipedia:Edit_warring}}, where opposing groups repeatedly override each other’s changes. \citet{schneider2010content} find that frequently edited or heavily viewed pages---a pair of characteristics that tend to co-occur~\citep{ratkiewicz2010characterizing}---show relatively frequent discussions on reversions or vandalism\footnote{Vandalism on wikis includes the insertion of hoaxes, offensive language, or nonsense text. Reversions are edits used to restore content to its previous, stable version.}. These findings suggest that article popularity and controversy are strongly linked, with heightened attention often coinciding with editorial instability.

To address such challenges, the Wikipedia community has established a comprehensive set of governance mechanisms aimed at mitigating conflict and preserving content quality. These include the well-known ``three-revert rule,'' temporary or indefinite page protection, editorial tagging to flag controversial material, and administrative measures such as user warnings, temporary suspensions, or permanent bans for persistently disruptive behavior.\footnote{See \url{https://en.wikipedia.org/wiki/Wikipedia:Edit_warring} for policy details.}

Research in this area has largely focused on editorial behaviors and article-level indicators of controversy. Various approaches have been developed to quantify editorial conflict on Wikipedia. For instance, one of the earliest and most widely used metrics is the count of reverts and total number of edits, where a high number of reverts suggests a contentious article~\citep{kittur2007he}. Another commonly used metric is the number of words deleted by one user from another's contributions. This approach defines disputes between user pairs based on the amount of content removed by each party from the other's edits~\citep{vuong2008ranking}. A third approach involves analyzing mutual reverts between two editors. When two contributors repeatedly undo each other's changes, it often signals the presence of a sustained disagreement~\citep{suh2007us}. Temporal features have also been employed in measuring controversy. One such measure is the inverse of the time interval between consecutive edits made by different users. Shorter intervals between conflicting edits may reflect intense editorial disputes~\citep{brandes2008visual}. Similarly, \citet{west2010detecting} and\citet{adler2011wikipedia} utilized temporal patterns of editing behavior to detect instances of vandalism on Wikipedia. Their central argument is that edits deemed problematic tend to be reverted more rapidly than non-problematic ones, suggesting that the time interval between an edit and its subsequent reversion can serve as a reliable indicator of disruptive activity. \citet{yasseri2014most} proposed a composite metric of controversy, denoted as $M$, which aggregates the weights of mutually reverting editor pairs and multiplies this sum by the total number of editors engaged with the article. This method captures both the depth and breadth of conflict within a Wikipedia page.

Although these methods provide valuable insights into the dynamics of editorial behavior and have been instrumental in identifying controversial content, they primarily focus on article-level interactions such as editing frequency, reversion patterns, and word-level deletions. The role of references and citations, particularly scientific ones, has been relatively understudied in this body of work. Given Wikipedia’s emphasis on verifiability and reliable sourcing---especially in articles on contentious topics---citations themselves can become a site of disagreement. Understanding how scientific publications are cited, disputed, or removed during Wikipedia conflicts could offer new perspectives on the intersection between scientific knowledge and public discourse.

\subsection{Scientific Citations and Disputes in Wikipedia}

Scientific citations serve not only to acknowledge prior research but also function as instruments of persuasion, authority, and boundary-setting within both academic and public discourse~\citep{nigel1977referencing, cozzens1989citations}. In contested contexts, citations are often used strategically to support specific claims, challenge opposing arguments, or align with particular epistemic communities~\citep{gilbert1984opening, hyland1999academic}.

On Wikipedia, where editorial guidelines such as ``verifiability'' and ``no original research'' place strong emphasis on reliable sourcing, scientific references play a particularly central role~\citep{benjakob2022citation, lewoniewski2023understanding, konieczny2016teaching}. Editors rely on scientific publications not only to substantiate statements, but also to resolve disputes, justify inclusion or exclusion of content, and negotiate editorial consensus. In articles on contentious topics, such as climate change, COVID-19, or gender identity, the selection and interpretation of specific scientific sources frequently becomes a focal point of disagreement~\citep{esteves2019anthropogenic, benjakob2022citation, currie2012feminist}. Despite the importance of citations in these disputes, the characteristics of the cited literature and how these characteristics may influence the likelihood of citation-related disputes are not yet well understood.

One potentially important factor that has received limited attention is the role of accessibility to scientific publications. OA publications, which are openly accessible to the public, are generally associated with increased visibility, broader dissemination, and greater uptake in social media and public discourse compared to subscription-based articles~\citep{piwowar2018state, bjork2012open}. This broader exposure may contribute to a higher likelihood of being cited on platforms like Wikipedia~\citep{yang2024open}. Because OA publications are more readily accessible to editors and readers alike, they may be more frequently included in articles, more actively scrutinized by the community, or more easily drawn into points of contention. However, whether OA articles are more likely to be involved in citation-related disputes remains an open empirical question.

Previous research has provided important insights into how scientific knowledge circulates in digital environments. Yet, few studies have directly examined how specific attributes of scientific publications---such as OA status, citation impact, disciplinary field, or publication venue---are associated with their presence in Wikipedia disputes. Moreover, most existing work has examined either editorial behavior or citation patterns in isolation, without integrating these perspectives into a unified empirical framework.

This study seeks to address these gaps by investigating how features of scientific publications, particularly OA status, are related to their involvement in citation disputes on Wikipedia. By combining large-scale Wikipedia editing histories with detailed bibliometric metadata from sources such as Crossref and OpenAlex, this research provides new insights into how scientific literature challenged and contested within one of the world’s most widely used open knowledge platforms.

\section{Methodology}
\label{sec:Methodology}

\subsection{Data Collection and Sources}

This study draws on data from three primary sources. First, we used Crossref Event Data curated by CWTS, which records citation-related events on Wikipedia. These events capture the addition and removal of references to scientific publications identified by Digital Object Identifiers (DOIs). Second, we utilized the Wikipedia API to retrieve article revision histories, editor identifiers, and page-level metadata such as namespace classification. Third, we integrated metadata from the OpenAlex dataset to obtain attributes of the cited scientific publications, including OA status, publication year, citation count, disciplinary classification, and retraction status.

\subsection{Temporal Scope, Dispute Definition, and Filtering}

Crossref Event Data has recorded citation events from April 11, 2017 onward. To ensure the completeness of revision histories from article inception, we restricted our sample to English-language Wikipedia articles whose earliest revision date occurred on or after January 1, 2018. This yielded 609,256 Wikipedia articles with a combined total of over 71 million revisions.

Previous studies on disputes in Wikipedia have primarily focused on edit wars, and adopt the definition provided directly by Wikipedia\footnote{ \url{https://en.wikipedia.org/wiki/Wikipedia:Edit_warring}}. These studies have centered on detecting such conflicts, typically using one of two approaches. The first involves analyzing the associated talk pages\footnote{ \url{https://en.wikipedia.org/wiki/Help:Talk_pages}} for indicators such as cleanup tags, strong language, or hostile interactions. In some cases, the length of the talk page may exceed that of the article itself, requiring the archiving of earlier discussions. The second approach examines the revision history of an article, where edit war behavior can often be identified by patterns of repeated reverts among contributors~\citep{sumi2011edit}.

In a broader sense, scientific disputes have been characterized as sustained and public disagreements involving conflicting knowledge claims. According to \citeauthor{mcmullin1987scientific}, a scientific controversy exists when both sides assert the validity of their own views while challenging the other's, and when these views are presented as being grounded in scientific reasoning~\citep{mcmullin1987scientific}. The disagreement must also be visible to a wider audience through written or spoken communication, rather than remaining private, allowing others to evaluate the competing claims. 

Building on these prior perspectives and considering the specific context of scientific publications, we operationalize a scientific dispute as a sequence of citation changes involving repeated addition and removal of the same DOI by at least two distinct editors within a short time frame. Specifically, a dispute is identified when the same DOI is inserted or removed at least three times within a seven-day period. 

We evaluated alternative temporal windows of one, three, seven, and thirty days. The seven-day window was selected as a compromise: shorter intervals risked excluding slower-developing disputes, while longer periods tended to capture unrelated content updates, obsolecense or page evolution. Applying this definition, and restricting the scope to article pages\footnote{We limit our analysis to Wikipedia pages that have a namespace of 0, which represent articles. A Wikipedia namespace is a categorization system that groups pages by their function or content type, helping distinguish between articles, user pages, discussions, drafts, and project-related content. \url{https://en.wikipedia.org/wiki/Wikipedia:Namespace}}, we identified 193,416 candidate scientific disputes across 2,807 Wikipedia pages and 9,602 unique DOIs.

\subsection{Dispute Validation and Consolidation}

An initial manual inspection of 100 randomly sampled events revealed that only 33 percent represented genuine disputes. The remaining events primarily stemmed from incomplete or incorrect records in the Crossref Event Data. To improve reliability, we implemented a two-step verification process.

First, for each event, we used the Wikimedia API to retrieve the article text before and after the corresponding revision. We then compared different versions of the same article and excluded events where the detected text differences did not include the corresponding DOI\footnote{This points to some data quality issues in the collection of DOIs in the Crossref Event Data.}. After filtering for actual DOI changes, we again randomly sampled 100 events for manual validation. Among these, 88 percent were confirmed as genuine scientific disputes. The remaining cases typically involved actual DOI changes, but lacked clear evidence of disagreement, such as edit summaries indicating a dispute. The random sampling was designed to ensure representativeness of the filtered dataset.

\begin{figure}[H]
    \centering
    \includegraphics[width=0.8\linewidth]{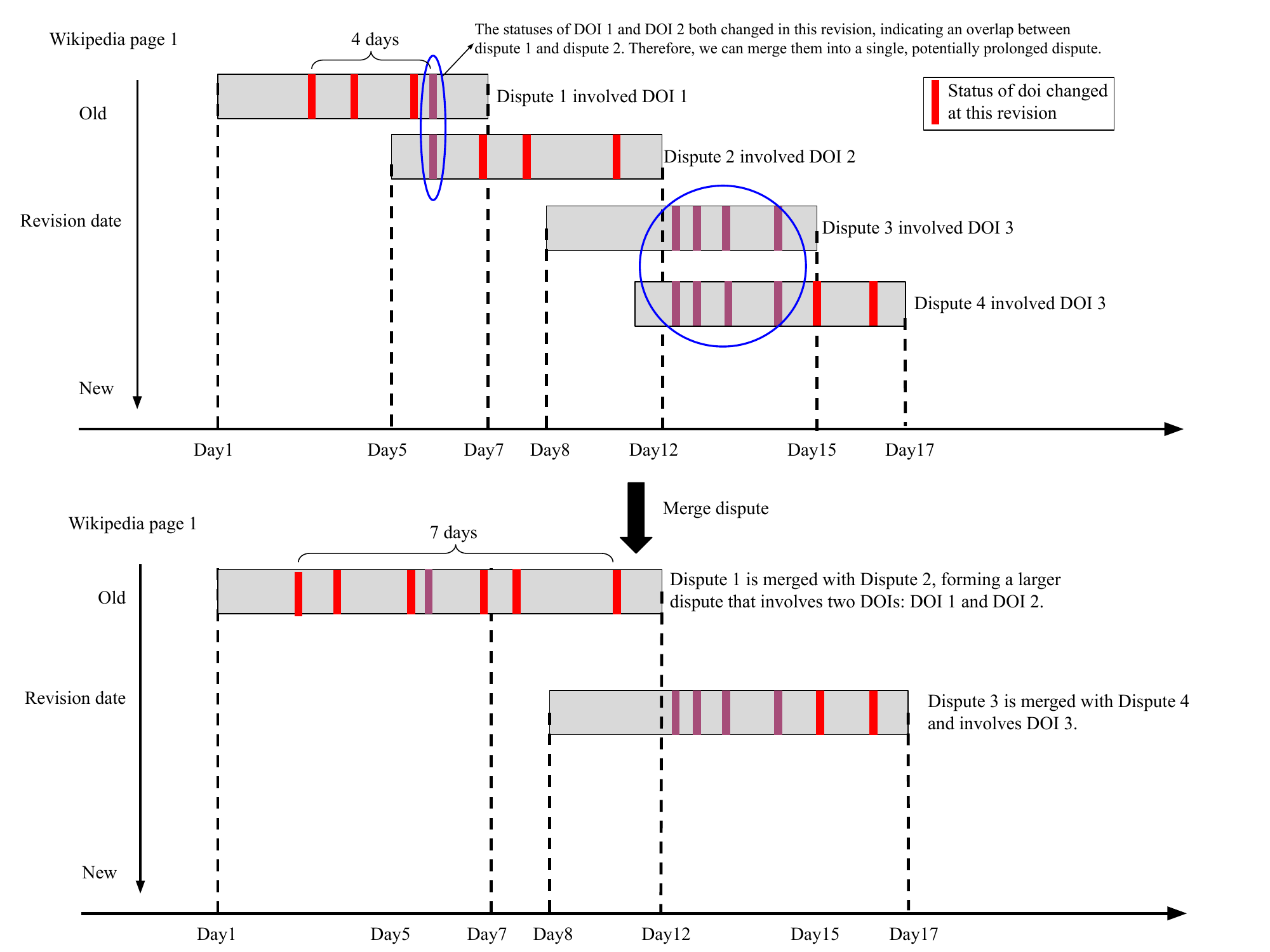}
    \caption{Merging procedure for scientific disputes.}
    \label{fig:Merge_process}
\end{figure}

After validation, we retained 10,884 dispute events involving 1,681 Wikipedia articles and 3,548 distinct scientific publications. To account for prolonged or fragmented disputes, we consolidated overlapping events on the same article into unified conflict episodes. Specifically, two disputes were merged if they occurred within overlapping time windows and involved at least one common revision, the process is illustrated in \ref{fig:Merge_process}. This conservative merging criterion minimized the risk of erroneously combining unrelated events, helping ensure that each episode reflected a coherent editorial disagreement over one or more scientific sources. The final consolidation process yielded 2,221 distinct scientific dispute episodes.

\subsection{Metadata Enrichment and Final Dataset}

To contextualize the dispute episodes, we enriched the dataset with structured metadata from both OpenAlex and Wikipedia API. For each cited scientific publication, we collected information on OA status, publication year, citation volume, primary disciplinary concept, and retraction status. Additionally, we extracted topic classifications for each Wikipedia article using the ORES ArticleTopic\footnote{\url{https://www.mediawiki.org/wiki/ORES/Articletopic}} model and retrieved structural characteristics of the articles.

After removing entries with incomplete metadata, the final dataset consisted of 2,221 consolidated scientific dispute episodes. These episodes involved 1,669 Wikipedia articles and 3,514 unique scientific publications, all with complete metadata coverage.

\section{Results}

\subsection{Characteristics of Publications Involved in Scientific Disputes}

\begin{figure}[H]
    \centering
    \includegraphics[width=0.8\linewidth]{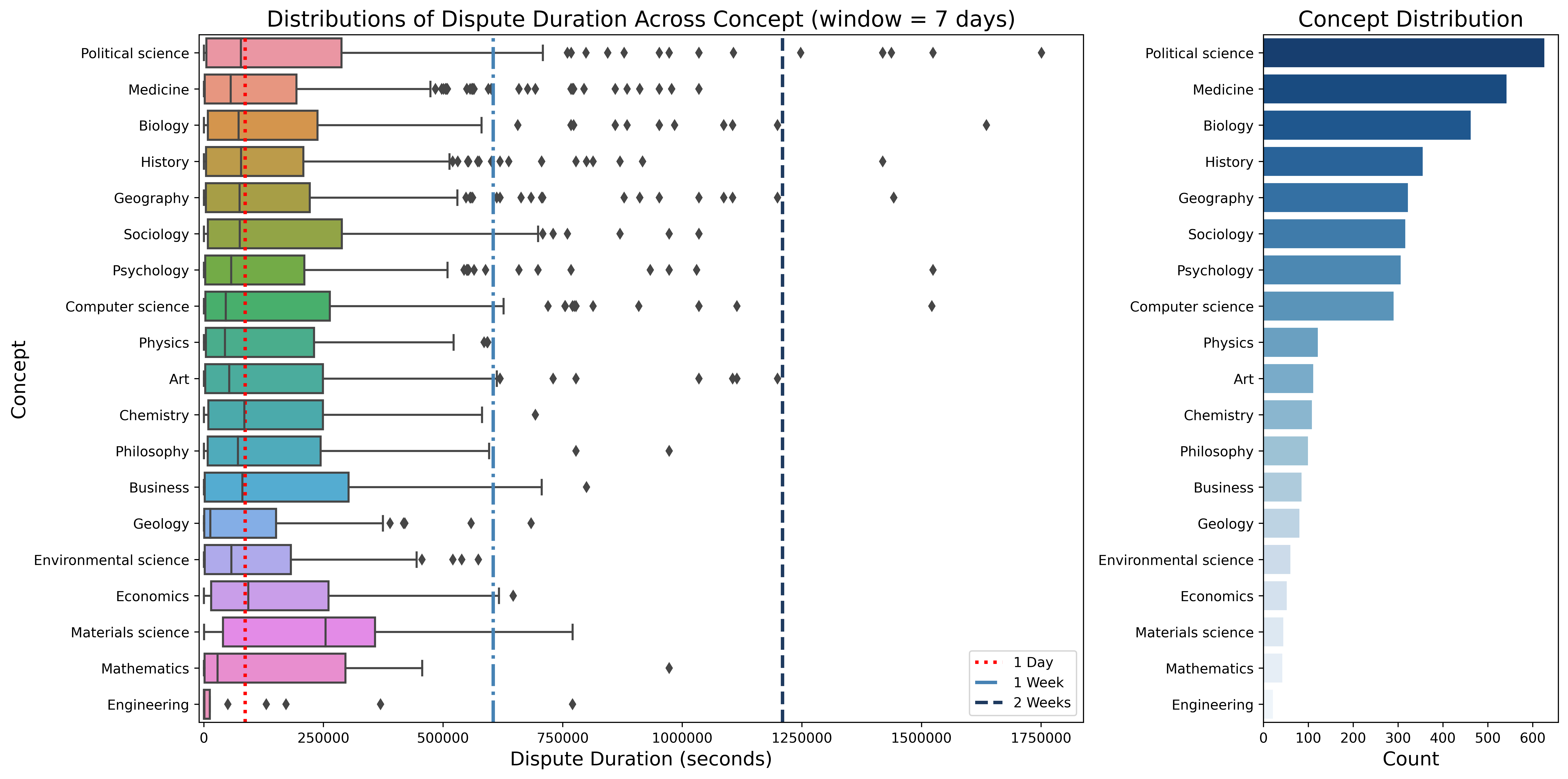}
    \caption{Distribution of scientific dispute durations across disciplines.}
    \label{fig:Distributions of Dispute Duration Across Fields}
\end{figure}

We begin by examining the distribution of scientific disputes across academic disciplines. Figure~\ref{fig:Distributions of Dispute Duration Across Fields} presents the frequency and duration of disputes for each primary concept, based on scientific articles. Concepts are ordered by the total number of dispute episodes (shown on the right), while the left panel displays the distribution of dispute durations in seconds. To aid interpretation, three vertical reference lines are included: a red dotted line marking one day, a light blue line marking one week, and a dark blue line indicating two weeks.

The most frequently contested fields include Political Science, Medicine, and humanities-related disciplines such as History, Geography, and Sociology. In contrast, foundational sciences like Physics, Chemistry, Mathematics, and Engineering are markedly less represented. Mathematics and Engineering show the lowest dispute frequencies.

In terms of duration, most disputes are short-lived, typically resolving within three days. The average dispute lasts less than one day, suggesting that many editorial disagreements are addressed or lose momentum quickly. Among the exceptions, materials science exhibits the longest average dispute durations, whereas geology shows the shortest. The relatively brief duration in Engineering may reflect the small number of disputes observed (only 21 articles). Overall, we find limited variation in dispute duration across disciplines, indicating that Wikipedia editors tend to resolve disputes at similar speeds regardless of topic. Alternatively, it is possible that unresolved disputes migrate to talk pages rather than continuing on the main article pages.

\begin{figure}[H]
    \centering
    \includegraphics[width=0.8\linewidth]{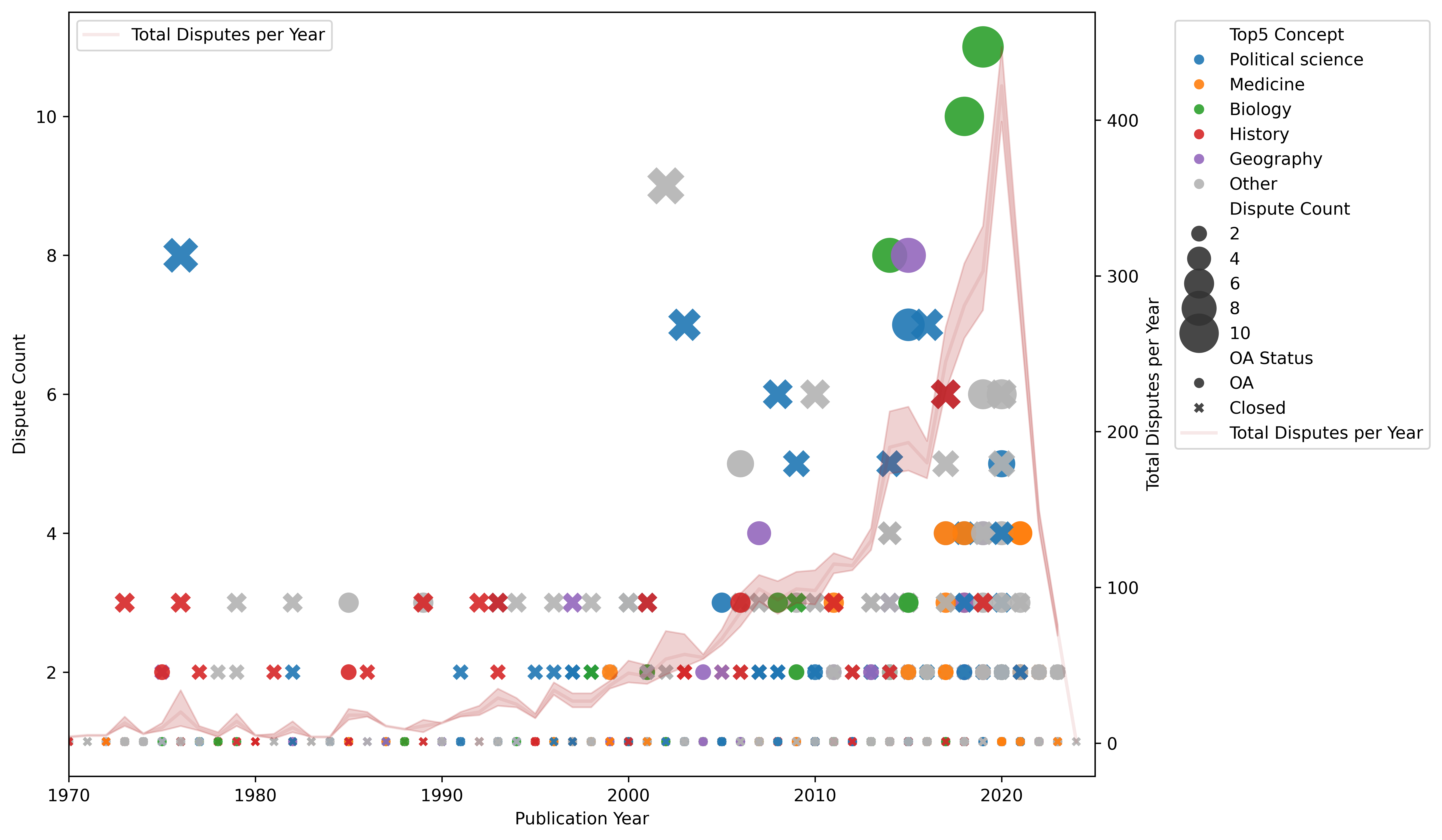}
    \caption{Temporal trends and disciplinary distribution of scientific disputes.}
    \label{fig:disputes&discipline}
\end{figure}

To further understand the temporal dynamics of scientific disputes, Figure~\ref{fig:disputes&discipline} shows when disputes occurred relative to publication year. Each point represents a scientific article involved in a dispute, with the x-axis showing publication year and the left y-axis representing dispute count. The point shape indicates access type (circles for OA, crosses for paywalled), while the point size corresponds to the number of dispute episodes per article. The red trend line, plotted against the right y-axis, represents the total number of disputes per year with a shaded 95\% confidence interval. For visual clarity, we color only the top 5 most disputed concepts; all others are grouped under ``Other'' in light gray. 

Most articles appear in only a single dispute episode (Figure~\ref{fig:disputes&discipline}). However, since the early 2000s, an increasing number of articles have been involved in multiple disputes. The article with the highest dispute frequency is a 2019 publication in Frontiers in Genetics titled ``Assortative Mating on Ancestry-Variant Traits in Admixed Latin American Populations'', which was cited in 11 separate dispute episodes. These disputes revolved around contested interpretations of genetic admixture and its implications for demographic classification and racial identity. Editors contested the reliability of the study, juxtaposing it with sources such as the CIA World Factbook and the Latinobarómetro survey. Disagreements involved content addition/removal, accusations of bias or vandalism, and broader tensions over the role of scientific sources in politically sensitive topics.

Overall, articles in Political Science and Biology are more frequently involved in high-frequency disputes. In contrast, articles in History, Sociology, and Psychology tend to be associated with more moderate or low-intensity dispute activity. Furthermore, we observe a growing number of disputes over time, especially concerning recently published articles, which are more likely to be cited and edited.

Next, we assess whether citation impact differs between disputed and non-disputed publications cited by Wikipedia. Using log-transformed citation counts, we find that disputed articles exhibit slightly higher citation levels (median $= 3.66$, mean $= 3.69$, SD $= 1.81$) than non-disputed ones (median $= 3.56$, mean $= 3.49$, SD $= 1.74$). A Mann–Whitney U test indicates that this difference is statistically significant ($p = 0.0002$), though modest in effect size. These findings suggest that articles involved in disputes tend to be marginally more cited, potentially reflecting their broader visibility or relevance to contested topics.

\begin{figure}[H]
    \centering
    \includegraphics[width=0.8\linewidth]{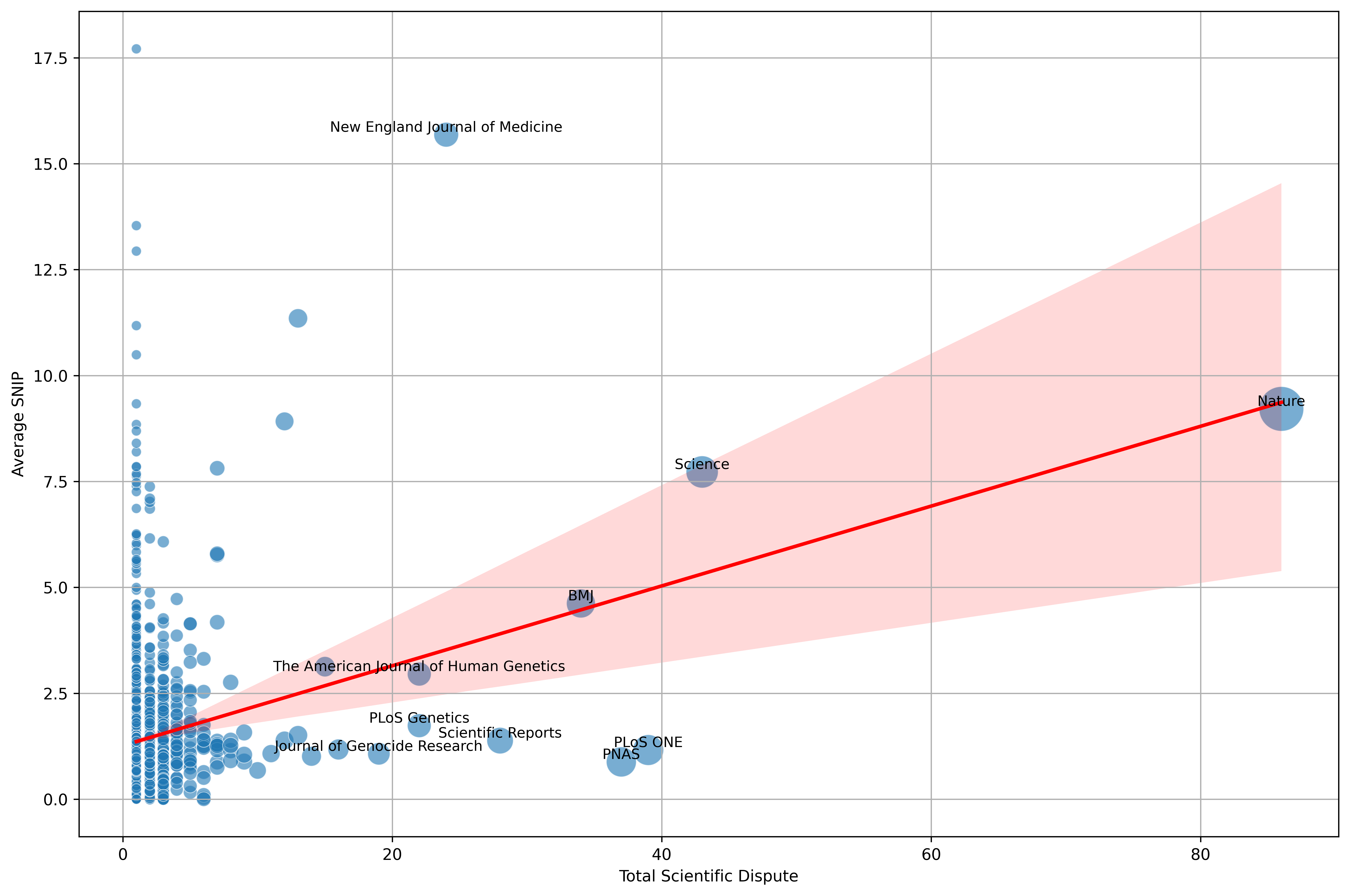}
    \caption{Dispute frequency and journal Prestige (SNIP).}
    \label{fig:journal_and_dispute}
\end{figure}

We also examine the role of journal prestige by comparing the top 10 journals most frequently associated with disputed publications. Figure~\ref{fig:journal_and_dispute} shows each journal's total number of disputes and corresponding average SNIP (source normalized impact per paper) scores\footnote{SNIP is a key metric from CWTS that reflects the average citation impact of a journal's publications.\url{https://www.journalindicators.com/}}. Highly prestigious journals such as Nature, Science, and the New England Journal of Medicine exhibit both high dispute frequencies and high SNIP scores. This suggests that articles from high-impact journals may be more likely to attract editorial scrutiny on Wikipedia.

Notably, some journals, such as PLoS ONE and PNAS, also rank high in dispute frequency despite relatively modest SNIP scores. These cases may reflect their multidisciplinary scope, high publication volumes, or frequent citation across diverse Wikipedia articles. This pattern suggests that both journal prestige and publication exposure may influence the likelihood of scientific disputes.

\begin{figure}[H]
    \centering
    \includegraphics[width=0.8\linewidth]{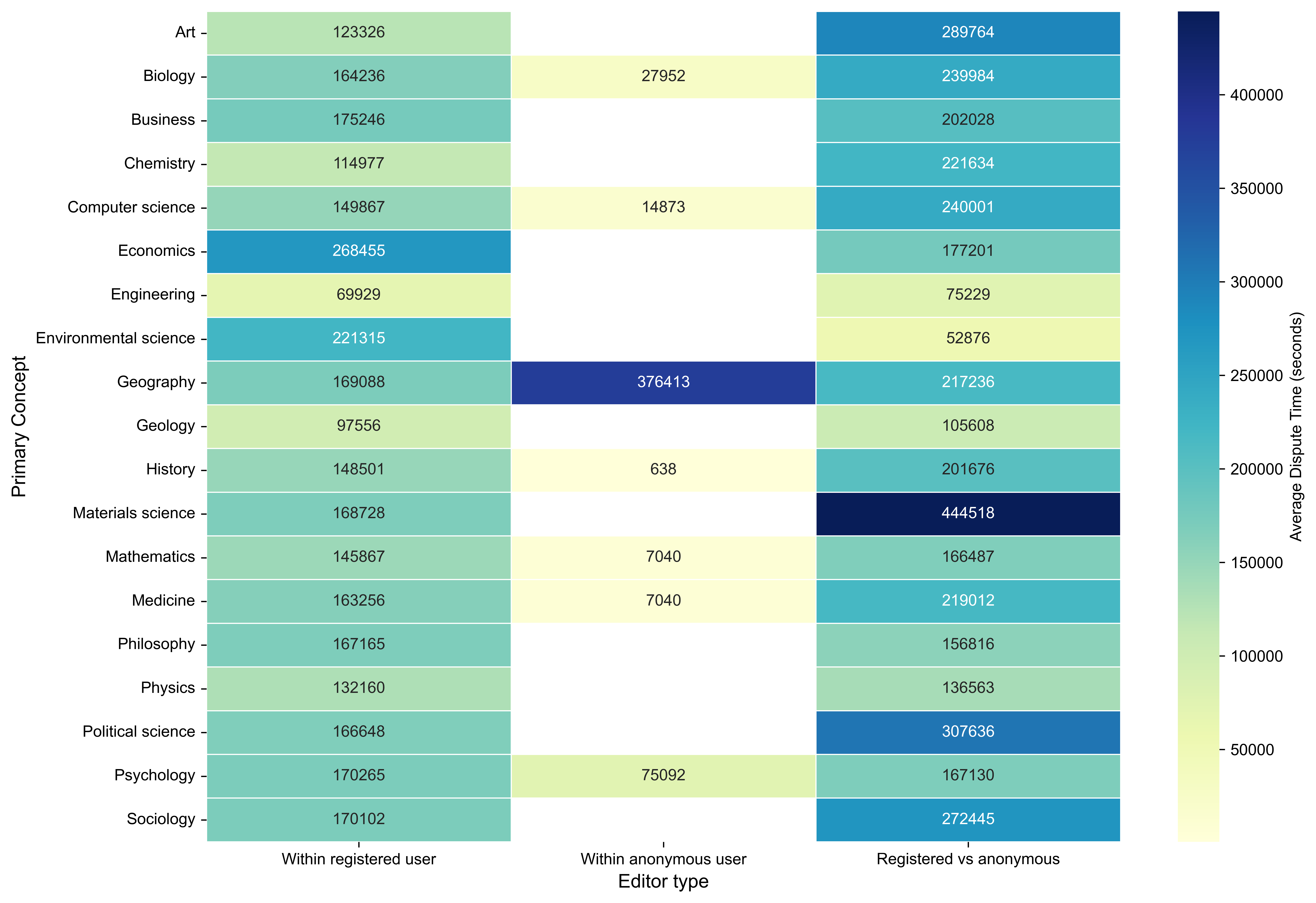}
    \caption{Average dispute duration by editor type and concept.}
    \label{fig:editor_type&concept&dispute_time}
\end{figure}

In addition, we distinguished between registered and anonymous editors by identifying whether an editor’s username was an IP address. Based on this classification, we analyzed the duration of scientific disputes across different primary concepts and editor types, as shown in Figure~\ref{fig:editor_type&concept&dispute_time}.

Each dispute was categorized into one of three editor-type groups: disputes exclusively involving registered editors, those involving only anonymous editors, and those jointly edited by both types. The y-axis in the figure represents different scientific concepts, while each cell in the heatmap indicates the average duration (in seconds) of disputes for the corresponding editor type and concept combination.

The results reveal several noteworthy patterns. First, the majority of scientific disputes occur either among registered editors or between registered and anonymous editors. In contrast, disputes occurring solely between anonymous editors are relatively rare and tend to be short-lived. An exception is observed in the field of Geography, where anonymous-only disputes lasted around four days on average.

Moreover, disputes involving both registered and anonymous editors generally lasted longer than those limited to registered editors, particularly in domains such as Materials Science, Political Science, Art, and Sociology. This suggests that dispute dynamics are not only influenced by the scientific topic at hand, but also by the type of contributors involved in the editing process. The presence of both registered and anonymous editors in a dispute may indicate more persistent disagreements or challenges in reaching consensus.

\subsection{Descriptive Analysis of Open Access Articles and Disputes}

To address the second research question regarding whether OA publications are more likely to be involved in scientific disputes, we begin with a descriptive overview.


First, we examined the distribution of OA types among the disputed articles in our dataset. Paywalled publications account for the largest share, comprising 54.3\% of all disputed articles. This pattern is consistent with earlier findings ~\citep{yang2024open} that indicate approximately 55\% of Wikipedia-cited scientific articles are paywalled. Among OA categories, the most frequent are Bronze (14\%), followed by Green (11.5\%), Gold (10.5\%), Hybrid (7.5\%), and Diamond (2.2\%).

\begin{figure}[H]
    \centering
    \includegraphics[width=0.8\linewidth]{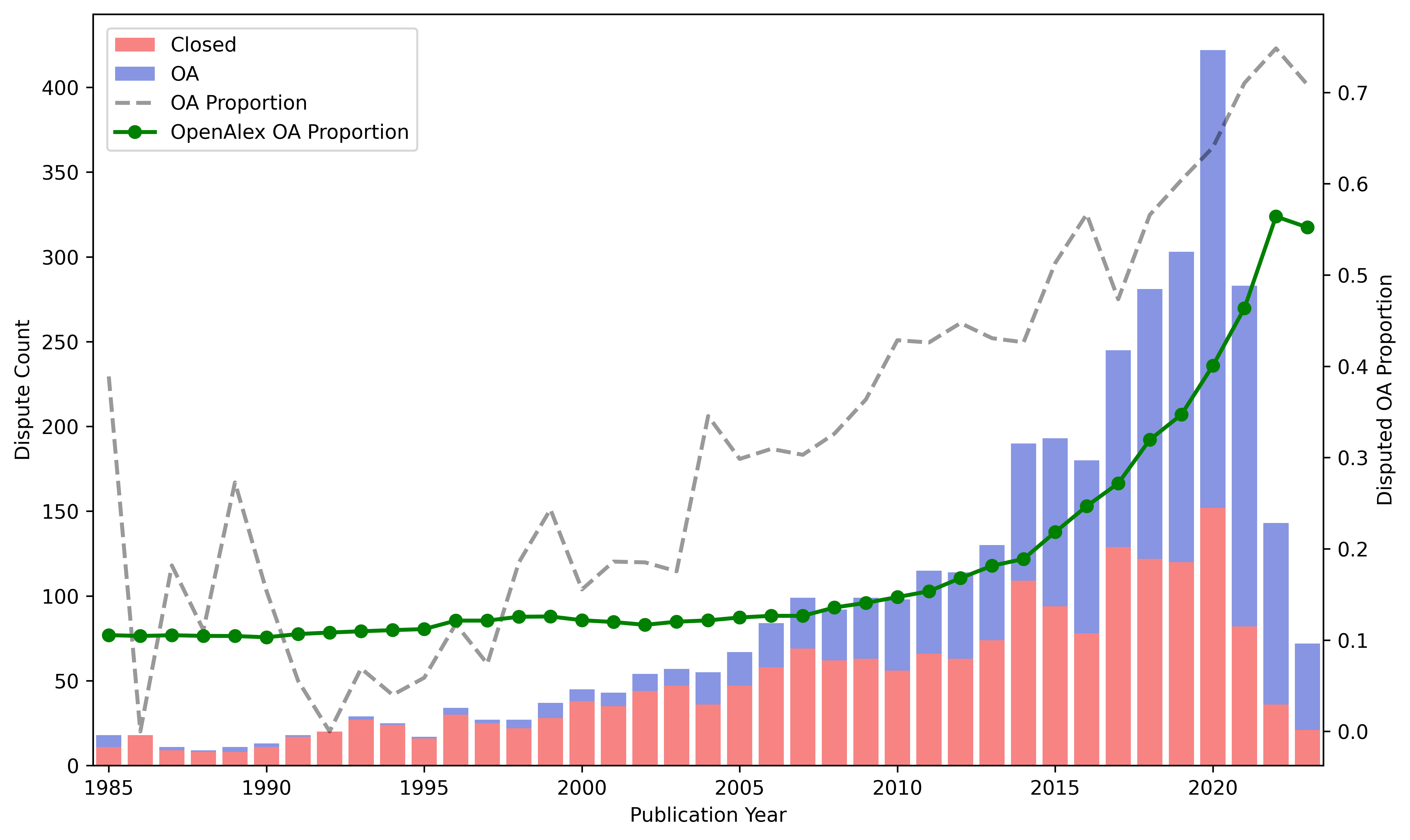}
    \caption{Temporal trends in OA proportion among disputed articles.}
    \label{fig:distribution of oa by time}
\end{figure}

Next, we examine how the proportion of OA articles involved in disputes has evolved over time (Figure~\ref{fig:distribution of oa by time}). Figure~\ref{fig:distribution of oa by time} shows a stacked bar chart based on publication year, with the left y-axis indicating the count of disputed articles and the right y-axis showing the proportion of OA articles. Closed articles are shown in red, and OA articles are in blue. The grey dotted line represents the annual proportion of OA articles among disputed publications, while the solid green line depicts the overall OA publication rate in the OpenAlex dataset.

First, we find that more recently published articles tend to be involved in a higher number of disputes. Second, the proportion of OA articles among disputed publications has increased steadily over time. This proportion often exceeds the overall OA prevalence observed in the broader scientific literature, particularly after the year 2000. These patterns suggest that OA articles may play an increasingly visible role in the contestation of knowledge on Wikipedia.

\begin{figure}[H]
    \centering
    \includegraphics[width=0.8\linewidth]{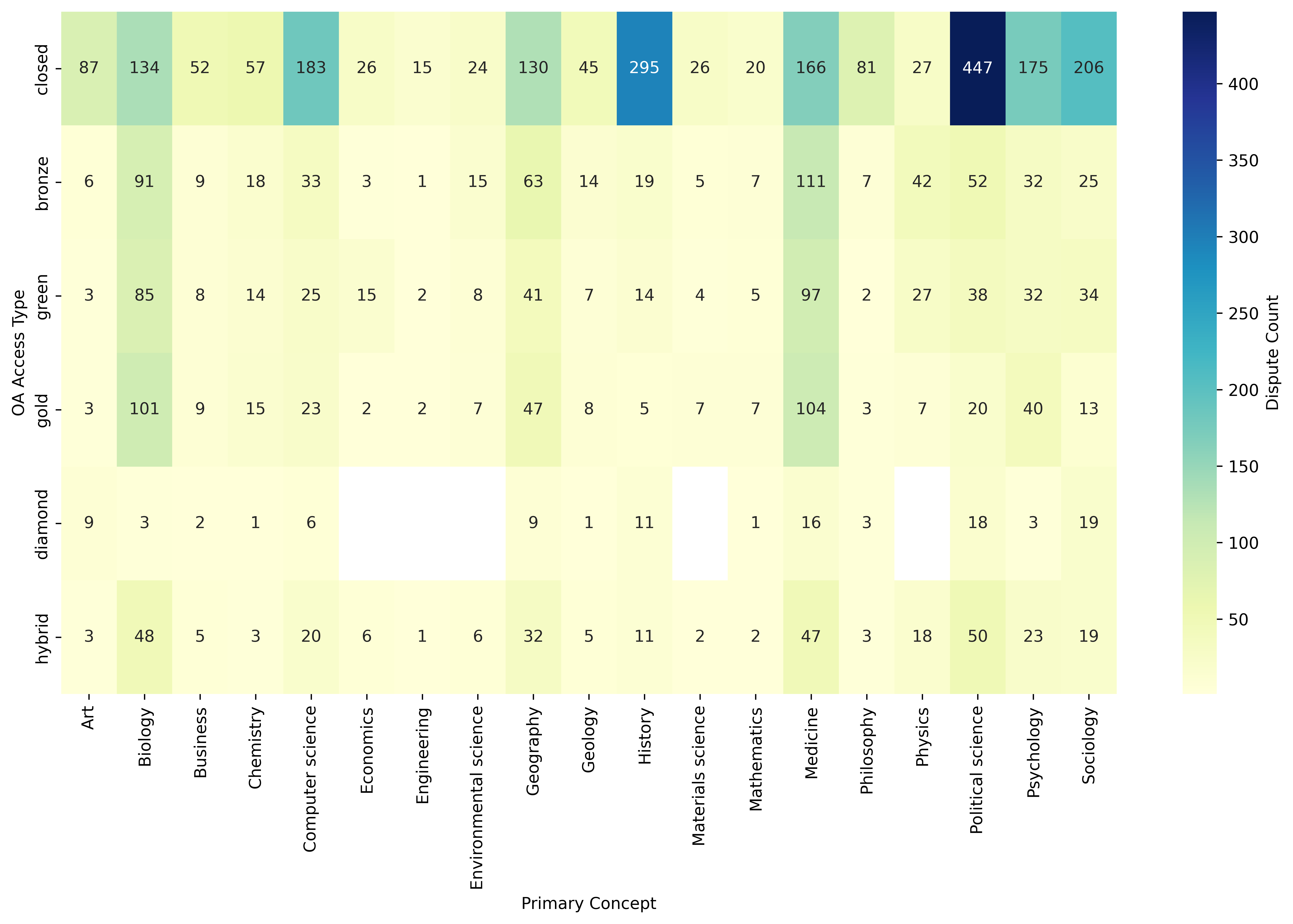}
    \caption{Heatmap of dispute frequency by OA type and concept.}
    \label{fig:Heatmap of Dispute Counts by OA Access Type and Concept}
\end{figure}

To explore disciplinary variation in OA involvement, we constructed a heatmap that illustrates the number of disputes involving each OA type across different scientific concepts (Figure~\ref{fig:Heatmap of Dispute Counts by OA Access Type and Concept}). Each cell reflects the frequency of disputes, with darker colors indicating higher counts.

The heatmap highlights clear differences across domains. Political Science exhibits the highest dispute frequency, with approximately 72\% of these disputes involving paywalled articles. In comparison, Biology and Medicine show a more balanced distribution across OA type. Although paywalled still dominates, the proportions of Green, Bronze, and Gold OA are relatively higher in these fields. Diamond OA appears least involved across all domains, which may reflect either lower publication volumes or different editorial practices.

These descriptive findings provide initial evidence that both OA status and disciplinary domain are associated with patterns of Wikipedia disputes involving scientific literature. Subsequent regression and survival analyses further test the statistical significance and interaction effects of these variables.

\subsection{The role of open access articles in scientific disputes}

To examine whether OA articles are more likely to become subjects of scientific disputes, we first classified our dataset into two groups: 3,514 articles identified as involved in scientific disputes, and all remaining articles cited in Wikipedia without dispute records (N = 1,140,242). A binary variable \texttt{has\_dispute} was created to indicate dispute involvement (1 = involved, 0 = not involved).

We constructed a logistic regression model with \texttt{has\_dispute} as the dependent variable. Key continuous predictors include:

\begin{itemize}
  \item \textbf{Article Age} (\texttt{ln\_article\_age}): The age of the article since publication, measured in months and log-transformed to reduce skewness.
  \item \textbf{Citation Count} (\texttt{ln1p\_cited\_by\_count}): The number of citations received, log-transformed after adding 1 to handle zero values.
  \item \textbf{OA Status} (\texttt{is\_oa}): A binary indicator where 1 represents OA articles and 0 represents paywalled articles.
  \item \textbf{Concept}: A categorical variable representing the subject area of the article. Dummy coding was applied, with \texttt{Engineering} used as the reference category.
  \item \textbf{Retraction Status} (\texttt{is\_retracted}): A binary indicator of whether the article has been retracted.
  \item \textbf{Number of References} (\texttt{n\_refs}): The total number of references cited by the article.
  \item \textbf{Topic}: A categorical variable representing the topic of the article, extracted from OpenAlex. Dummy coding was applied, with \texttt{Modeling the Dynamics of COVID-19 Pandemic} used as the reference category.
\end{itemize}


\begin{table}[htbp]
\centering
\caption{Logistic regression predicting likelihood of Wikipedia dispute involvement.}
\begin{tabular}{lccc}
\toprule
\textbf{Variable} & \textbf{Coefficient} & \textbf{\textit{p}-value} & \textbf{Odds Ratio} \\
\midrule
Intercept & -4.069 & $<$0.001 & 0.017 \\
OA (is\_oa) & 0.891 & $<$0.001 & 2.438 \\
Log Article Age (ln\_article\_age) & -0.521 & $<$0.001 & 0.594 \\
OA $\times$ Log Article Age & -0.212 & $<$0.001 & 0.809 \\
Log Citation Count (ln1p\_cited\_by\_count) & 0.197 & $<$0.001 & 1.218 \\
\# References (n\_refs) & -0.0017 & $<$0.001 & 0.998 \\
Retracted (is\_retracted) & 0.562 & 0.429 & 1.754 \\
\addlinespace
Art & 0.427 & 0.062 & 1.532 \\
Biology & -0.465 & 0.038 & 0.628 \\
Business & 0.247 & 0.307 & 1.280 \\
Chemistry & -0.310 & 0.165 & 0.734 \\
Computer Science & 0.145 & 0.511 & 1.156 \\
Economics & 0.284 & 0.226 & 1.328 \\
Environmental Science & 0.543 & 0.076 & 1.721 \\
Geography & 0.345 & 0.119 & 1.412 \\
Geology & -0.609 & 0.044 & 0.544 \\
History & 0.793 & 0.001 & 2.210 \\
Materials Science & -0.312 & 0.210 & 0.732 \\
Mathematics & -0.234 & 0.406 & 0.792 \\
Medicine & 0.271 & 0.218 & 1.311 \\
Philosophy & 0.645 & 0.008 & 1.906 \\
Physics & -0.054 & 0.821 & 0.947 \\
Political Science & 1.090 & $<$0.001 & 2.974 \\
Psychology & 0.464 & 0.056 & 1.591 \\
Sociology & 0.986 & $<$0.001 & 2.680 \\
\bottomrule
\end{tabular}
\vspace{0.5em}

\footnotesize{\textit{Note:} Reference category for disciplines is Engineering. Pseudo $R^2 = 0.0357$.}
\label{tab:logit_dispute}
\end{table}

Results (see Table~\ref{tab:logit_dispute}) indicate that OA articles are significantly more likely to be involved in disputes ($\beta = 0.891$, $p < 0.001$), suggesting that publicly accessible research is more frequently cited and scrutinized. Articles with higher citation counts also show a positive association with dispute likelihood ($\beta = 0.197$, $p < 0.001$), consistent with the idea that high-impact or widely discussed works attract more editorial attention. Conversely, older articles are less likely to be disputed ($\beta = -0.521$, $p < 0.001$), especially among OA publications, as indicated by a significant interaction term ($p < 0.001$).

Disciplinary differences are also evident. Compared to Engineering articles, those in Political Science ($\beta = 1.09$), Sociology ($\beta = 0.99$), History ($\beta = 0.79$), and Philosophy ($\beta = 0.64$) are significantly more likely to be disputed ($p < 0.01$ for all). 

We conducted a similar analysis using article topic instead of concept. 



We limited topics to the top 10 most frequently occurring among disputed articles, grouped all others under an ``Other'' category, and used the smallest of the top 10—\textit{Modeling the Dynamics of COVID-19 Pandemic}—as the reference.

\begin{table}[htbp]
\centering
\caption{Logistic regression predicting likelihood of Wikipedia dispute by topic.}
\begin{tabular}{lccc}
\toprule
\textbf{Variable} & \textbf{Coefficient} & \textbf{\textit{p}-value} & \textbf{Odds Ratio} \\
\midrule
Intercept & -2.092 & $<$0.001 & 0.124 \\
OA (is\_oa) & 0.603 & 0.005 & 1.828 \\
Log Article Age (ln\_article\_age) & -0.482 & $<$0.001 & 0.618 \\
OA $\times$ Log Article Age & -0.176 & $<$0.001 & 0.839 \\
Log Citation Count & 0.143 & $<$0.001 & 1.153 \\
\# References & -0.0021 & $<$0.001 & 0.998 \\
Retracted & 0.346 & 0.628 & 1.413 \\
\addlinespace
American Political Thought and History & -0.681 & 0.021 & 0.506 \\
Coronavirus Disease 2019 & 0.214 & 0.459 & 1.239 \\
COVID-19 Research & -0.008 & 0.974 & 0.992 \\
Genomic Analysis of Ancient DNA & 1.237 & $<$0.001 & 3.445 \\
Intersectionality in LGBTQ+ Mental Health & -0.008 & 0.977 & 0.992 \\
Other & -1.776 & $<$0.001 & 0.169 \\
Islamic Reform in Middle East & -0.115 & 0.672 & 0.892 \\
Populism in Contemporary Politics & 0.744 & 0.012 & 2.104 \\
Stellar Astrophysics & -1.598 & $<$0.001 & 0.202 \\
Politics in Turkey & 0.622 & 0.030 & 1.863 \\
\bottomrule
\end{tabular}
\vspace{0.5em}

\footnotesize{\textit{Note:} Reference topic is ``Modeling the Dynamics of COVID-19 Pandemic''. Pseudo $R^2 = 0.04195$.}
\label{tab:logit_topic}
\end{table}

Again, OA status is positively associated with dispute involvement (OR = 1.83, \(p = 0.005\), see see Table~\ref{tab:logit_topic}). Newer articles are more likely to be disputed, particularly when they are OA, as indicated by the significant interaction effect between OA status and article age. Citation count is also positively associated with dispute likelihood, whereas the number of references has a weak negative association. Retraction status does not significantly predict dispute involvement.

At the topic level, articles categorized under \textit{Genomic Analysis of Ancient DNA} exhibit significantly higher odds of dispute (OR = 3.44, \(p < 0.001\)), potentially reflecting ethical or interpretive controversies. Topics related to contemporary politics—such as \textit{Populism in Contemporary Politics} and \textit{Politics in Turkey}—are also associated with higher dispute likelihoods. Conversely, technical fields like \textit{Stellar Astrophysics} show significantly lower odds of disputes, suggesting that such topics are less contentious in the public or editorial sphere. Articles grouped under ``Other'' topics are the least likely to be disputed, further confirming that less socially relevant or debated topics are less prone to editorial contention.

\begin{figure}[H]
    \centering
    \includegraphics[width=0.8\linewidth]{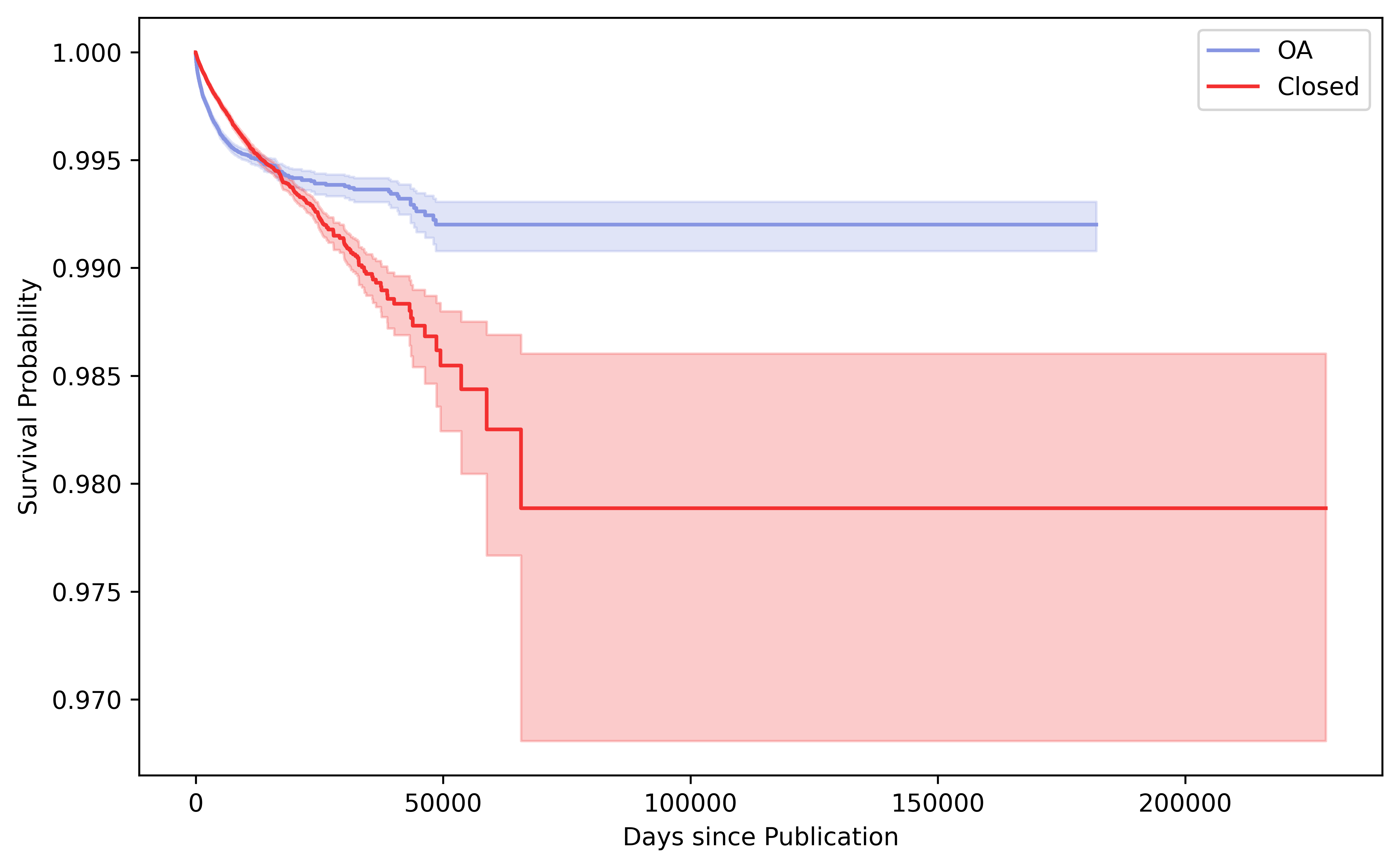}
    \caption{Kaplan-Meier survival curves for time to first dispute by OA status.}
    \label{fig:survival_line}
\end{figure}

While logistic regression provides insight into the likelihood of dispute involvement, we were also interested in understanding how quickly OA articles become subject to disputes compared to paywalled articles. We defined survival time as the number of days between an article’s publication and its first involvement in a dispute. We then applied Kaplan-Meier survival analysis, stratified by OA status, to visualize the time-to-dispute distributions, as shown in Figure~\ref{fig:survival_line}.

In the figure, the x-axis represents days since publication, and the y-axis indicates survival probability (i.e., the probability of not having been involved in a dispute). Blue and red lines correspond to OA and paywalled articles, respectively, with shaded areas indicating 95\% confidence intervals. 

The survival curves reveal notable differences between the two groups. In the early stage after publication, OA articles tend to become involved in disputes more quickly than paywalled articles, as indicated by the steeper, near-exponential decline of the OA curve compared to the more gradual, linear-like decrease of the paywalled curve. At later time points, however, this pattern reverses: the OA curve stabilizes,  whereas the paywalled curve continues to decline, reflecting a more persistent accumulation of disputes over time. A log-rank test confirms that the difference is statistically significant (\(p < 7.6 \times 10^{-15}\)).

It should also be noted that the majority of articles in the dataset (approximately one million) never become involved in disputes, while only about 3,000 do. This substantial imbalance explains the narrow confidence intervals observed across most of the curves, alongside the much wider intervals in the long-tail region, where data are sparse.

To further explore the timing of dispute involvement, we fitted a Cox proportional hazards model using OA status, citation count, and concept as predictors. Results are shown in Table~\ref{tab:cox_results}.

\begin{table}[htbp]
\centering
\caption{Cox Proportional Hazards Regression Results}
\label{tab:cox_results}
\begin{tabular}{lrrrr}
\toprule
\textbf{Variable} & \textbf{Coef} & \textbf{HR (exp(coef))} & \textbf{p-value} & \textbf{Significance} \\
\midrule
OA (is\_oa) & 0.37 & 1.45 & <0.005 & *** \\
Citation count (log, ln1p\_cited\_by\_count) & 0.11 & 1.11 & <0.005 & *** \\
Art & -0.31 & 0.73 & 0.20 &  \\
Biology & 0.10 & 1.11 & 0.65 &  \\
Business & 0.56 & 1.75 & 0.02 & * \\
Chemistry & -1.24 & 0.29 & <0.005 & *** \\
Computer science & -0.03 & 0.97 & 0.90 &  \\
Economics & -0.43 & 0.65 & 0.09 &  \\
Environmental science & 1.67 & 5.29 & <0.005 & *** \\
Geography & 0.21 & 1.23 & 0.35 &  \\
Geology & 0.61 & 1.84 & 0.01 & ** \\
History & 1.92 & 6.79 & <0.005 & *** \\
Materials science & -0.60 & 0.55 & 0.02 & * \\
Mathematics & -0.39 & 0.68 & 0.15 &  \\
Medicine & 0.77 & 2.15 & <0.005 & *** \\
Philosophy & 0.62 & 1.86 & 0.01 & ** \\
Physics & 0.03 & 1.03 & 0.89 &  \\
Political science & 2.37 & 10.72 & <0.005 & *** \\
Psychology & 1.80 & 6.07 & <0.005 & *** \\
Sociology & 0.27 & 1.32 & 0.22 &  \\
\bottomrule
\end{tabular}
\begin{tablenotes}
\small
\item \textit{Note:} Reference category for discipline is \textbf{Engineering}. Concordance: 0.60; Significance levels: *** $p<0.005$, ** $p<0.01$, * $p<0.05$.
\end{tablenotes}
\end{table}

The analysis reveals that OA articles are significantly more likely to be disputed sooner (HR = 1.45, \(p < 0.005\)). Citation count is slightly positively associated with dispute timing (HR = 1.11, \(p < 0.005\)), suggesting that highly cited papers may attract disputes more quickly, possibly because they draw greater attention and scrutiny. Relative to Engineering, articles in Chemistry and Materials Science are significantly less likely to be disputed promptly. In contrast, articles in Environmental Science, Geology, History, Medicine, Philosophy, Political Science, and Psychology exhibit substantially higher hazards of dispute. This disciplinary variation may reflect differences in public visibility, societal relevance, and the extent to which research in these areas intersects with contentious or value-laden issues.

Lastly, we also investigated whether scientific disputes involving OA publications tend to reach consensus more quickly, as reflected by shorter dispute durations, compared to disputes involving paywalled publications. Using the Mann–Whitney U test, we found no statistically significant difference in dispute duration between OA and paywalled articles (p = 0.274). Additionally, a multivariate regression analysis controlling for primary concept, citation count, and article age showed that OA status was not a significant predictor of dispute duration ($\beta$ = 0.1215, p = 0.213).

\section{Discussion}

This study systematically analyzed the characteristics of scientific disputes on Wikipedia and examined the role of OA publications in such disputes. The discussion is organized around the two research questions (RQs) and grounded in the empirical findings presented in the Results section.

Regarding \textbf{RQ1}: \textit{What are the characteristics of scientific publications that are involved in scientific disputes?} Our findings show that scientific disputes occur more frequently in the social sciences and humanities, particularly in fields such as Political Science, History, and Sociology. These disciplines often engage with issues related to social values, ideological differences, and political viewpoints, which are associated with higher epistemic uncertainty and interpretative flexibility. In contrast, fields such as the natural sciences and engineering, which tend to rely on more empirically stable knowledge, are associated with fewer citation-related disputes.

Although the focus of our study differs, this disciplinary pattern aligns with prior research by \citet{yasseri2014most}, who found that the most controversial articles on English Wikipedia often concern topics such as individuals, politics, religion, nations, and global warming. A similar pattern was observed in their cross-linguistic study of the 1,000 most controversial Wikipedia articles across ten language editions~\citep{sumi2011edit}. While these articles are not necessarily based on scientific publications, they reveal a thematic overlap with some fields in our dispute dataset.

To explore this further, we compared our list of scientific disputes with the M-score of article-level controversy as defined by \citet{sumi2011edit}. The results indicate little overlap between highly controversial Wikipedia articles and those containing numerous scientific disputes. For example, the English Wikipedia article for ``Jesus,'' which ranks among the top 10 most controversial with an M-score of 8,421,728, contains only five scientific disputes. This divergence suggests that public attention to controversial topics does not necessarily translate into contention over scientific sources. Scientific disputes differ in that they require a different type of engagement, rooted in interpretive legitimacy and epistemic trust.

Furthermore, the resolution time for scientific disputes is relatively short. Across nearly all Wikipedia concepts in our dataset (with the exception of materials science), the median time to resolution is less than one day. This may reflect the perception of scientific publications as authoritative sources, which facilitates faster consensus-building. Although prior work has not provided exact durations, agent-based modeling studies suggest that reaching consensus typically requires more than 20 rounds of edits~\citep{kalyanasundaram2015agent}. Additionally, the presence of heterogeneous editor types (e.g., registered and anonymous users) has been associated with longer conflict duration~\citep{kalyanasundaram2015agent}. Our findings support this observation: disputes involving both registered and anonymous editors generally last longer than those involving a single user type. One possible explanation is that anonymous contributors are often viewed as less credible, and previous studies have shown that edits by anonymous users are more likely to be reverted and contain disruptive content~\citep{tran2020anonymity}.

We found that disputed articles tend to have higher citation counts and are frequently published in prestigious journals such as \textit{Nature}, \textit{Science}, and \textit{PNAS}. These journals are also among the most cited sources on Wikipedia overall~\citep{arroyo2020science, yang2022map}. This aligns with previous findings that scientific publications from high-status journals tend to be more visible to both editors and readers~\citep{teplitskiy2017amplifying}. Their broad reach and authoritative status increase the likelihood of being cited but also raise the potential for contention when used as evidence in public knowledge spaces like Wikipedia.

Regarding \textbf{RQ2}: \textit{Are OA publications more likely to be used in scientific disputes compared to paywalled publications?} Our analysis shows that OA publications are significantly more likely to be involved in disputes, and that disputes concerning OA articles tend to emerge sooner after publication. This pattern can be explained by the increased accessibility of OA articles~\citep{yang2024open}, which facilitates rapid dissemination and critical evaluation by a broader and more diverse editor base. These results extend existing literature on OA’s role in enhancing visibility, readership, and scholarly engagement~\citep{yang2024open}. Furthermore, OA articles may disproportionately address emerging, interdisciplinary, or controversial topics, which naturally attract greater contestation and discussion. This is consistent with prior findings that disputes around contested content on Wikipedia often mirror broader societal controversies~\citep{borra2015societal}.

The steady increase in the proportion of disputed OA articles, outpacing the general growth of OA publishing, suggests that openness not only increases accessibility but also accelerates the dynamics of knowledge contestation. These patterns support the idea that OA fosters a more participatory and transparent knowledge ecosystem, where scientific claims are more rapidly scrutinized and negotiated. However, the relationship between OA and disputes varies across disciplines. For example, in Political Science, disputed articles more frequently involve paywalled publications, possibly reflecting disciplinary norms regarding information gatekeeping or the prominence of certain traditional sources~\citep{severin2020discipline}. In biological and medical sciences, the distribution of OA and paywalled articles in disputes is more balanced, indicating different editorial cultures and accessibility patterns. These disciplinary differences suggest that openness interacts with field-specific epistemic cultures and norms, reinforcing the idea that the impact of OA on knowledge contestation is context-dependent.

Although previous research suggests that non-experts may endorse scientific norms more strongly in highly contested fields~\citep{schug2025endorsement}, potentially facilitating consensus-building, our findings do not support the notion that open access significantly shortens the duration of disputes. This suggests that while openness may enhance transparency and participation, dispute resolution is likely shaped more by the nature of contested content and the dynamics of editorial deliberation than by access alone.

Taken together, these findings point to several implications. The higher involvement of OA publications in disputes indicates that OA plays a critical role in enabling public scrutiny and facilitating dynamic knowledge negotiation in collaborative platforms. This reinforces the broader vision of OA not only as a means for democratizing access to scientific information but also as a catalyst for enhanced critical engagement. Concurrently, the prevalence of disputes in social sciences and humanities underscores the ongoing challenges of reconciling diverse perspectives in fields where social, political, and ideological values are deeply embedded.

Despite these contributions, this study has limitations. First, we relied on the Crossref Event Data, which, while valuable for efficiently capturing citation events on Wikipedia, has limitations in temporal coverage and completeness due to discrepancies with Wikipedia’s own data records. Although strict filtering ensured data quality, some relevant dispute events may have been missed, potentially biasing results toward more recent or well-documented cases. Second, our operationalization of scientific disputes, defined by a seven-day time window, while justified through sensitivity analyses, may benefit from further refinement and validation in future work. Third, as our analysis focused solely on Wikipedia’s main article pages and did not account for activity on associated talk pages~\citep{arroyo2024citation}, we may have underestimated the duration or complexity of some disputes. Future research should incorporate talk page discussions to more comprehensively understand how disputes involving scientific articles emerge, evolve, and are eventually resolved. Finally, since disputes may manifest differently across language editions of Wikipedia~\citep{yasseri2014most}, and our study only examined the English version, further work is needed to explore cross-lingual variations in the dynamics and representation of scientific disputes.

In summary, this study demonstrated that scientific disputes on Wikipedia are shaped by the interplay of disciplinary context, publication impact, editor identity, and openness. OA publications emerge as prominent actors in these disputes, facilitating earlier and more frequent contestation. These insights contribute to a deeper understanding of how scientific knowledge is constructed and contested in an open, collaborative digital environment and highlight the complex role of openness in shaping knowledge dynamics.

\section{Conclusion}

This study investigated the characteristics of scientific publications involved in disputes on Wikipedia and examined the role of OA publications in these conflicts. Our findings reveal that scientific disputes are more prevalent in social sciences and humanities, where epistemic uncertainty and value-laden topics are common, compared to the natural sciences and engineering. Publications that attract greater attention, such as those with high citation counts and those published in prestigious journals, are more likely to be contested. Registered editors play a central role in sustaining these disputes, indicating the importance of experienced contributors in the editorial process.

Furthermore, OA publications are significantly more likely to be used in scientific disputes and tend to become subjects of dispute earlier than paywalled publications. This suggests that increased accessibility and visibility foster more rapid and frequent contestation of scientific knowledge. Nevertheless, the relationship between OA and disputes varies across disciplines, highlighting the influence of field-specific norms on the dynamics of knowledge negotiation.

These findings highlight a dual role of OA: while it democratizes access to scientific knowledge, it also intensifies public scrutiny and epistemic contestation. Wikipedia serves as a key arena where scientific authority is continuously negotiated, not only through content but through the strategic deployment of citations. Our results offer implications for open science, scholarly communication, and digital platform governance, suggesting that the architecture of openness affects not only access, but also the trajectories of scientific controversy in public discourse.

\bibliographystyle{unsrtnat}
\bibliography{references}  






\end{document}